\documentstyle[preprint,aps]{revtex}
\begin{document}

\draft

\title{Effect of Quantum Confinement on Electron
       Tunneling through a Quantum Dot}

\author{Kicheon Kang$^{a,b}$ and B. I. Min$^{a,c}$ }

\address{ $^a$Department of Physics,
          Pohang University of Science and Technology\\
          Pohang 790-784, Korea\\
          $^b$Department of Physics,
          Korea University \\ Seoul 136-701, Korea \\
          $^c$Max-Planck-Institut f\"{u}r Festk\"{o}rperforschung,\\
                D-70506 Stuttgart, Germany }

\date{\today}

\maketitle

\begin{abstract}
Employing the Anderson impurity model, we study tunneling properties 
through an ideal quantum dot near the conductance minima.
Considering the Coulomb blockade and the quantum confinement
on an equal footing, we have obtained current contributions 
from various types of tunneling processes;
inelastic cotunneling, elastic cotunneling, and resonant
tunneling of thermally activated electrons. We have found that
the inelastic cotunneling is suppressed in the quantum confinement
limit, and thus the conductance near its minima is determined by the elastic
cotunneling at low temperature ($k_BT \ll \Gamma$, $\Gamma$: 
dot-reservoir coupling constant), or by the resonant tunneling of single
electrons at high temperature ($k_BT \gg \Gamma$).
 
\end{abstract}
%
During last decade, there has been a rapid advance in the field 
of single electronics, and accordingly much scientific attention has been
given to transport properties through ultra-small tunnel junctions 
such as GaAs quantum dot\cite{kastner92,kastner93,korotkov96}.  
In a quantum dot with small capacitance,
``Coulomb blockade" of tunneling occurs for small bias voltage $V$
when the charging energy (Coulomb energy $U$) in the dot
is sufficiently large as compared to a thermal energy $k_BT$.
It occurs because even a single tunneling event increases the electrostatic
Coulomb energy of the system considerably.
However, even in this regime, a finite current
can flow via virtual intermediate states
arising from the quantum fluctuation of macroscopic electric
charge in the central electrode of the system.
This process in a quantum dot, so called cotunneling or
macroscopic-quantum-tunneling,
was first pointed out by Averin and coworkers\cite{averin}, and is
considered as setting a limit to the performance accuracy 
of the single electron transistor.
They have shown that the transport near conductance minima is dominated
by the inelastic cotunneling process
involving the creation of an electron-hole excitation
in the central electrode,
and predicted an algebraic variation of the leakage current
with applied voltage ($\sim V^3$) and temperature ($\sim T^2$).
The theory of cotunneling has been derived within the lowest 
order perturbation when the energy discreteness in a quantum dot
is not important, {\it i.e.} the continuous energy spectrum is
assumed in the central electrode.
The inelastic cotunneling has been observed both in metal-insulator-metal
tunnel junctions\cite{geerligs90,eiles92} and in a 2D electron system of
GaAs/Ga(Al)As heterostructure\cite{glattli91,pasquier93}.

When dealing with an ultra-small quantum dot, the effect of level discreteness 
(energy quantization: $\Delta$) becomes very important.
The effect will be more prominent 
in semiconductor systems than in metallic systems, due to
much lower electron concentration ($\rho$) and lower effective electron 
mass ($m^*$) in semiconductor systems (recall that 
a free electron approximation yields 
$\Delta=\frac{1}{g(\varepsilon_F) V} = \frac{ 2\hbar^2\pi^2 }
{ Vm^* (3\pi^2\rho)^{1/3} }$ for 3D systems).
There have been quite a few experimental evidences exhibiting
coexistence of the charge and energy quantization in the tunneling
properties \cite{korotkov96}.
Furthermore, it is shown that the level spacing $\Delta$ can be even 
comparable to the Coulomb energy $U$ for a Si-based quantum dot
transistor \cite{leobandung95}. For such
systems, the ``quantum confinement" will become significant as much as the 
Coulomb blockade for the tunneling in the single electron transistor.
This kind of electronic confinement could be realized for an ultra-small
dot at relatively high temperature. 
Illustrating typical parameters, a Si dot with diameter of 20nm 
would have $\Delta\sim 25$ meV, 
and then the confinement of electron could be realized even at $T <{\cal O}(100)$K 
for $U\sim 15$ meV \cite{leobandung95}.

We address in this paper whether the inelastic cotunneling
phenomenon is really a limiting factor in operating 
single electron quantum dot devices. 
For this purpose, we have examined tunneling properties of
an ideal quantum dot coupled to two reservoirs in terms of
the Anderson impurity model, where the quantum confinement is
important as much as the Coulomb blockade. Special attention is
focused on the temperature dependence of the inelastic cotunneling
by treating the Coulomb blockade and the quantum confinement on
an equal footing.  We have found that the characteristic of the tunneling in
the quantum confined system far 
from conductance maxima is qualitatively different from the case 
where the level discreteness can be neglected. 
In this case, the ``inelastic" cotunneling is substantially
suppressed at low temperature,
while the ``elastic" cotunneling or the resonant single electron tunneling 
of thermally activated electrons dominates in the system.
Therefore it is expected that, in small enough quantum dots,
there would be no substantial limitation on the performance accuracy in
practical devices by a macroscopic quantum tunneling of charge.

We start with the simplest Anderson model Hamiltonian\cite{anderson61}
to describe an ideal
quantum dot (labeled by $D$) weakly coupled to two electron reservoirs 
(labeled by $L$ and $R$):
\begin{eqnarray}
 {\cal H} &=& {\cal H}_L+{\cal H}_R+{\cal H}_D+{\cal H}_T  \nonumber \\
 {\cal H}_{L(R)} &=& \sum_{k,\alpha \in L(R)}\varepsilon_k
                 c^{\dagger}_{k\alpha}c_{k\alpha}  \\
 {\cal H}_D &=& \sum_{\alpha} \varepsilon_{\alpha} d^{\dagger}_{\alpha}
                d_{\alpha} + U\sum_{\alpha>\alpha'}n_{\alpha}n_{\alpha'} 
                \nonumber \\
 {\cal H}_T &=& \sum_{k,\alpha \in L,R} \left( V_{k\alpha}c^{\dagger}_{k\alpha}
                d_{\alpha} + \mbox{h.c.} \right) . \nonumber
\end{eqnarray}
Here ${\cal H}_L, {\cal H}_R, {\cal H}_D$ and ${\cal H}_T$ represent 
Hamiltonians of the left reservoir, the right reservoir, an interacting dot,
and tunneling between the dot and reservoirs, respectively. 
The levels in the dot are labeled
by an index $\alpha$, and $U$ denotes the Coulomb interaction between
electrons in the dot. The states in the reservoirs with energies
$\varepsilon_k$ are coupled to the dot by hopping matrix
element $V_{k\alpha}$.
The transition rate of an electron between the level $\varepsilon_{\alpha}$
and the reservoir $L(R)$ is given by
\begin{equation}
 \Gamma_{L(R)}^{\alpha}(\omega) = 2\pi \sum_{k \in L(R)} \left| V_{k\alpha}
    \right|^2 \delta(\omega-\varepsilon_k) .
\end{equation}
This Anderson-type Hamiltonian has been 
recently employed to describe transport in the quantum dot 
structure\cite{ng88,meir91,hersh91,meir93}. 
In the Coulomb blockade and the quantum confinement limit, we have
\begin{equation}
 U,\Delta \gg k_BT, \Gamma_{L(R)}, eV  
\end{equation}
where $eV$ corresponds to a potential energy difference coming from the
bias voltage across two reservoirs.

The cotunneling refers to a simultaneous tunneling of two electrons
through intermediate states with an extra electron or hole in a quantum dot. 
This second order cotunneling process is called ``inelastic" if there remains 
an electron-hole excitation after the process (Fig1.(a)), whereas the
process is said to be ``elastic" if no electron-hole excitation is left 
(Fig1.(b)).
The inelastic cotunneling current can be simply calculated by using the
Fermi golden rule. 

The initial eigenstate $|I\rangle$ of the Hamiltonian ${\cal H}_0={\cal H}_L
+{\cal H}_R+{\cal H}_D$ can be written as $|\phi_L\rangle |\psi^N\rangle
|\phi_R\rangle$, where $|\phi_{L(R)}\rangle$ denotes the Fermi sea of 
left (right) reservoirs, and $|\psi^N\rangle$ represents an $N$-particle
eigenstate of ${\cal H}_D$. There are two kinds of final states for inelastic
cotunneling processes. Those are $|F_1\rangle$ (an electron(hole) tunnels
from $L(R)$ to $R(L)$), and $|F_2\rangle$ (an electron(hole) tunnels
from $R(L)$ to $L(R)$), such that $|F_1\rangle=c^{\dagger}_{p\beta}d_{\beta}
d^{\dagger}_{\alpha}c_{k\alpha} |I\rangle$, $|F_2\rangle=c^{\dagger}_{k\alpha}
d_{\alpha}d^{\dagger}_{\beta}c_{p\beta} |I\rangle$ ($k\in L, p\in R$).
The inelastic cotunneling yields the current as 
\begin{equation}
 I^{in}(V) = e\left(\gamma_1(V)-\gamma_2(V)\right),
\end{equation}
where $\gamma_1$ and $\gamma_2$ are statistical averages of the
transition rates from the initial state $\{ |I\rangle \}$ to the final
states $\{ |F_1\rangle \}$ and $\{ |F_2\rangle \}$, respectively,
and are expressed as
\begin{equation}
 \gamma_i = \left\langle  \frac{2\pi}{\hbar} \sum_{F_i}
              |\langle I | {\cal H}_T \frac{ 1 }{ E_I-{\cal H}_0 } 
              {\cal H}_T | F_i \rangle |^2  \delta(E_I-E_{F_i})
            \right\rangle_I \;\; i=1,2 ,
\end{equation}
with $\langle\cdots\rangle_I$ denoting the statistical average over
the initial states.

For a constant $\Gamma_{L(R)}=
\Gamma_{L(R)}^{\alpha}(\varepsilon)$, $\gamma_1,\gamma_2$ are given by
\begin{mathletters}
 \label{eq:gamma}
\begin{eqnarray}
 \gamma_1(V) &=& \frac{ \Gamma_L\Gamma_R }{ h } \sum_{\alpha\ne\beta}
             \left\langle n_{\beta}(1-n_{\alpha}) \right\rangle
             \int d\varepsilon d\varepsilon' f(\varepsilon)
             (1-f(\varepsilon')) \nonumber \\
        & & \times \left(
             \frac{ 1 }{ \varepsilon_{\alpha}-\varepsilon+\mu_a }
           + \frac{ 1 }{ \varepsilon'-\varepsilon_{\beta}+\mu_b } 
                   \right)^2 
             \delta\left( \varepsilon'-\varepsilon_{\beta}+
              \varepsilon_{\alpha}-\varepsilon-eV \right) \\
 \gamma_2(V) &=& \gamma_1(-V) .  
\end{eqnarray}
\end{mathletters}
Here $\langle \cdots \rangle$ denotes statistical average,
and $f(\varepsilon)
=1/(e^{\beta\varepsilon}+1)$ is the Fermi distribution function for 
the reservoirs. $\mu_a(\mu_b)$ represents the charging energy of the
virtual intermediate state with one extra electron (hole). They are defined
by
\begin{mathletters}
\begin{eqnarray}
 \mu_a &=& NU-\mu_L , \\
 \mu_b &=& \mu_R-(N-1)U ,
\end{eqnarray}
\end{mathletters}
with $\mu_L(\mu_R)$ being the chemical potential of the left (right) 
reservoir which satisfies the relation $\mu_L-\mu_R=eV$. 
Equation (\ref{eq:gamma}) 
describes well the inelastic cotunneling  when the system is not too close 
to the conductance maxima.

Note that, if one assumes a continuous spectra
for electrons in the dot ($\Delta\rightarrow 0$),
Eq.(\ref{eq:gamma}) becomes equivalent to the result by
Averin and Nazarov \cite{averin}, 
since then $\left\langle n_{\beta}(1-n_{\alpha})
\right\rangle = f_{\beta}(1-f_{\alpha})$ with $f$ being the Fermi distribution 
function.  
However, in the quantum confinement limit, 
the inelastic cotunneling process described by Eq.(\ref{eq:gamma}) 
gives rise to drastically modified characteristics
due to following two reasons. 
First, for the electrons in the dot, the Fermi distribution function 
cannot be used in the quantum confinement limit ($k_BT,eV\ll \Delta$), 
because the number fluctuation is very small \cite{beenakker91,kubo92}. 
Second, more important difference results from the fact that there is 
no available excitation for inelastic cotunneling process with energy
$\varepsilon$ such that $0<\varepsilon<\Delta$.
Therefore it is expected that the inelastic cotunneling rate is much 
suppressed in the quantum confinement limit.

To see the functional form of the inelastic cotunneling current in the quantum 
confinement limit, it will be sufficient to consider lowest two 
levels. It is because contributions from other levels would be exponentially small
as compared to those of two levels. Let's assign the two levels as 
$\varepsilon_1=0$ and $\varepsilon_2=\Delta$ with $N=1$
for the ground state of the dot. Then we have $\langle n_1n_2\rangle=0$,
and Eq.(\ref{eq:gamma}) reduces to 
\begin{eqnarray}
 \gamma_1(V) &=& \frac{ \Gamma_L\Gamma_R }{ h } \int d\varepsilon\,
             f(\varepsilon) (1-f(\varepsilon'))  \nonumber \\
 & & \times \left[ \langle n_1 \rangle
            \left(
             \frac{ 1 }{ \Delta-\varepsilon +\mu_a }
           + \frac{ 1 }{ \varepsilon'+\mu_b }
                   \right)^2
             \delta\left( \varepsilon'-\varepsilon +\Delta-eV 
                   \right) \right. \label{eq:twolevel} \\
 & & + \left.\langle n_2 \rangle
            \left(
             \frac{ 1 }{ -\varepsilon +\mu_a }
           + \frac{ 1 }{ \varepsilon'-\Delta+\mu_b }
                   \right)^2
             \delta\left( \varepsilon'-\varepsilon -\Delta-eV 
                   \right) \right] . \nonumber 
\end{eqnarray}
Here $\langle n_1\rangle = 1/(1+e^{-\beta\Delta})$ and
$\langle n_2\rangle = e^{-\beta\Delta}/(1+e^{-\beta\Delta})$.
The first(second) term in Eq.(\ref{eq:twolevel}) describes the 
excitation (relaxation) process from initially ground (excited) state in the dot.
Further, in the quantum confinement limit, $\langle n_2\rangle$ is 
exponentially small, and so a contribution from the second term can be neglected.
Then using $\langle n_1\rangle \simeq 1$, Eq.(\ref{eq:twolevel}) becomes
\begin{equation}
 \gamma_1(V) \simeq \frac{ \Gamma_L\Gamma_R }{ h } 
    \int d\varepsilon\, f(\varepsilon)f(\Delta-eV-\varepsilon) 
         \left(
            \frac{ 1 }{ \Delta-\varepsilon +\mu_a }
           + \frac{ 1 }{ \varepsilon-\Delta+eV+\mu_b }
                   \right)^2 . \label{eq:gamma1}
\end{equation}
Since $f(\varepsilon)f(\Delta-eV-\varepsilon)$ is peaked 
around $\varepsilon\sim \Delta/2$, Eq. (\ref{eq:gamma1}) is approximated by
\begin{eqnarray}
 \gamma_1(V) &\simeq& \frac{ \Gamma_L\Gamma_R }{ h }
         \left( \frac{ 1 }{ \mu_a+\Delta/2 }
              + \frac{ 1 }{ \mu_b-\Delta/2 } \right)^2 
         \int d\varepsilon\,f(\varepsilon)f(\Delta-eV-\varepsilon) \label{eq:gamma1b} \\
  &\simeq& \frac{ \Gamma_L\Gamma_R }{ h }
         \left( \frac{ 1 }{ \mu_a+\Delta/2 }
              + \frac{ 1 }{ \mu_b-\Delta/2 } \right)^2 
          (\Delta-eV)\,e^{\beta eV}e^{-\beta\Delta} . \nonumber
\end{eqnarray}

Notable in  Eq. (\ref{eq:gamma1b}) is that
the inelastic cotunneling current in the quantum confinement limit
has a temperature dependence $e^{-\beta\Delta}$ suggesting
very much suppression at low temperature.
This is in contrast to $T^2$-dependence Averin and 
Nazarov have obtained for $\Delta\rightarrow0$ limit \cite{averin}.
The suppression of the inelastic cotunneling in the
quantum confinement limit originates from the reduction of 
phase space for the electron-hole excitations in the dot. Hence,
in this limit, other kinds of processes such as elastic cotunneling or 
single electron tunneling of thermally activated electrons is expected to
determine conductance minima\cite{glattli93}. 

The contributions from the elastic cotunneling and
the single electron tunneling of thermally activated electrons
can be estimated by Landauer-type formula 
through an interacting region \cite{meir92}. 
For the Anderson Hamiltonian, the conductance takes
the following form in the linear response regime,
\begin{equation}
 G = \frac{e^2}{\hbar}\tilde{\Gamma} \int d\varepsilon
     \left( -\frac{ \partial f }{ \partial\varepsilon }
     \right)
     \sum_{\alpha} \rho_{\alpha}(\varepsilon) , \label{eq:landauer}
\end{equation}
where $\tilde{\Gamma}=\frac{ \Gamma_L\Gamma_R }{ \Gamma_L+\Gamma_R }$,
$f(\varepsilon)=1/(e^{\beta(\varepsilon-\mu)}+1)$ $(\mu=\mu_L=\mu_R)$, and 
$\rho_{\alpha}(\varepsilon)=-\frac{1}{\pi} \mbox{Im}G_{\alpha}
(\varepsilon)$ is local density of states of electrons with label
$\alpha$. Equation of motion - decoupling method can be used to get
an approximate solution of the Green's function $G_{\alpha}$ 
\cite{meir91,kang95}. 
In the present study, more subtle Kondo effect will not be taken into 
account \cite{hersh91,meir93}.
Then Green's functions can be expressed as
\begin{equation}
 G_1(\varepsilon) \sim \frac{ 1 }{ \varepsilon+i\Gamma/2 }, \;\;\;
 G_2(\varepsilon) \sim \frac{ 1 }{ \varepsilon-(\Delta+U)+i\Gamma/2 }, 
\end{equation}
where $\Gamma=\Gamma_L+\Gamma_R$.

The conductance near its minima shows a different feature depending on
the relative size of $\Gamma$ and $k_BT$. For $\Gamma \gg k_BT$,  
$\left( -\frac{\partial f}{\partial\varepsilon} \right)$ 
in Eq. (\ref{eq:landauer}) can be approximated by
$\delta(\varepsilon-\mu)$ and the elastic cotunneling dominates (Fig1.(b)):
\begin{equation}
 G \simeq \frac{e^2\Gamma_L\Gamma_R}{ h } \left(
       \frac{1}{\mu^2} + \frac{ 1 }{ (\Delta+U-\mu)^2 }
                                    \right)  .
\end{equation}
On the other hand, for $\Gamma\ll k_BT$, there appears a limit in which 
the `resonant' tunneling of thermally activated electrons (Fig1.(c)) 
becomes most
important. In this case, $\rho_1(\varepsilon)\simeq 
\delta(\varepsilon), \rho_2(\varepsilon)\simeq\delta(\varepsilon-
\Delta-U)$, and thus we have
\begin{equation}
 G \simeq \frac{ e\tilde{\Gamma} }{ \hbar } \left[ -f'(0)
   -f'(\Delta+U)                        \right]  .
\end{equation}
In any case, the inelastic cotunneling process has much smaller contribution
to the conductance than other processes.
Present results can be summarized as in Fig. 2, which shows a schematic 
diagram of off-resonance tunneling processes.
We have examined tunneling  properties in the quantum confinement regime ($T/\Delta \ll 1$), 
in which the elastic cotunneling (A) or the resonant tunneling (B) is dominating.
Note that the inelastic cotunneling (C), which has been dealt with in 
previous studies \cite{averin}, becomes significant only in the 
opposite limit ($T/\Delta \gg 1$).

In conclusion, we have analyzed off-resonance tunneling properties
of an ideal quantum dot coupled to two reservoirs, in terms of 
the Anderson impurity model. We have found that, in the quantum confinement
limit, the inelastic cotunneling of electrons is much suppressed,
which is contrary to the case of previously studied $\Delta\rightarrow 0$ limit.
The suppression of the inelastic cotunneling current originates from 
the absence of phase space for the electron-hole excitations
in the dot with energy $\varepsilon$ such that $0 < \varepsilon < \Delta$. 
Hence, near the conductance minima, the transport is governed by the elastic 
cotunneling or by the resonant tunneling of thermally activated electrons.
The present result strongly suggests that the 
usually thought performance limitation of
single electron devices arising from the inelastic cotunneling phenomena
would be considerably reduced in an ultra-small quantum dot at 
sufficiently low temperature.

Acknowledgments$-$
The authors thank C.M.Ryu for his critical reading of our paper.
This work was supported by the POSTECH-BSRI program of the KME and 
the POSTECH special fund, and in part by the KOSEF-SRC program of
SNU-CTP. K.K. was partially supported by KOSEF- Post-Doc. program.
B.I.M. would like to thank L. Hedin and the MPI-FKF for the
hospitality during his stay. 


%
%
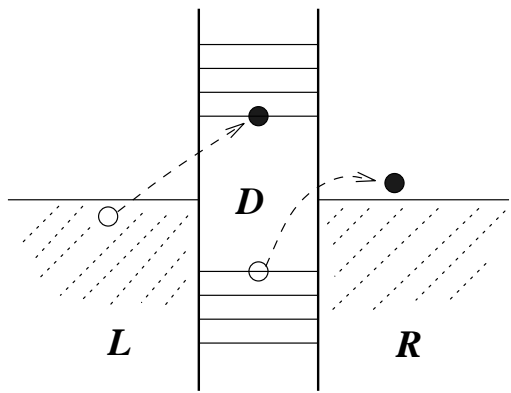
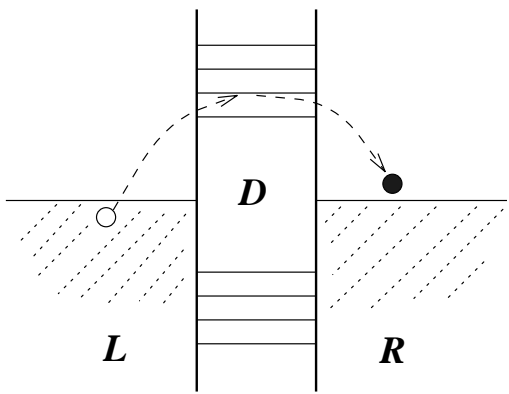
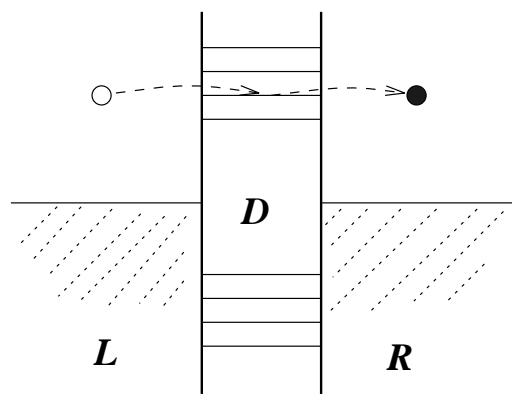
\begin{figure}
 \caption{ Various types of electron tunneling processes near conductance
  minima; (a) inelastic cotunneling, (b) elastic cotunneling, and (c)
  resonant tunneling of thermally activated electron. }
\end{figure}

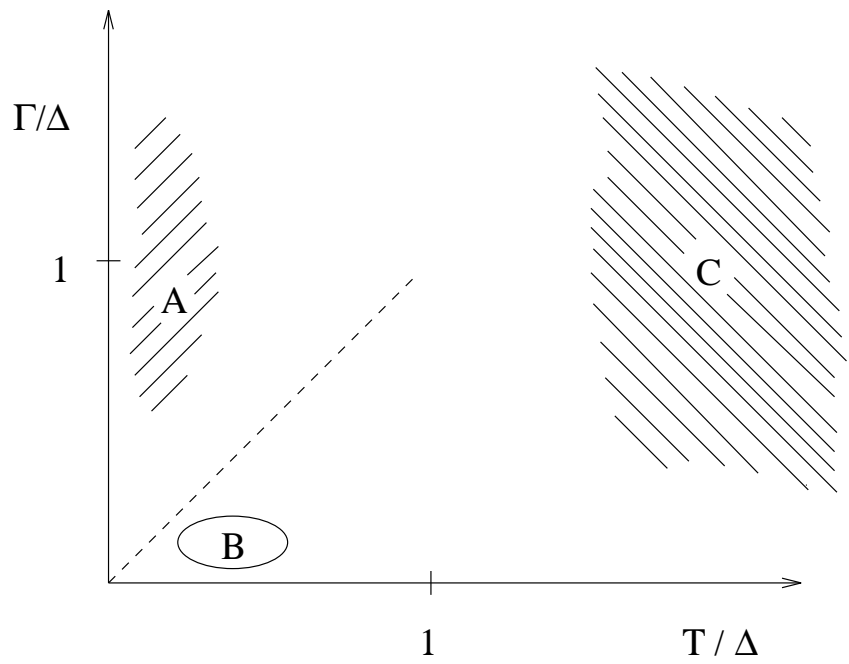
\begin{figure}
 \caption{Schematic diagram of showing various off-resonance tunneling processes 
    for different regimes defined by the ratios $T/\Delta$ and $\Gamma/\Delta$.
    A, B, and C correspond to regions in which tunneling processes of
    the elastic cotunneling, the resonant tunneling of activated 
    electrons, and the inelastic cotunneling are dominating, respectively. 
    The Coulomb blockade limit $U \gg kT$ is being considered here.}
\end{figure}

\end{document}